\RequirePackage{amsmath}
\documentclass[12pt]{iopart}
\usepackage{mathtools}
\usepackage{graphicx}
\usepackage{color}

\newcommand{\eps}{\epsilon}

\begin{document}

\title{A microscopic model of ballistic-diffusive crossover}
\author{Debarshee  Bagchi$^{1}$  and P. K. Mohanty$^{1,2}$}
\ead{pk.mohanty@saha.ac.in}
\address{  
 $^{1}$Condensed Matter Physics Division, Saha Institute of Nuclear Physics,1/AF Bidhan Nagar, Kolkata 700064, India\\
 $^{1,2}$Max Planck Institute for the Physics of Complex Systems, N\"othnitzer Straße 38, 01187 Dresden, Germany
 }
%\date{\today}

\begin{abstract}Several low-dimensional systems show a crossover from diffusive to ballistic heat
transport when system size is decreased. Although there is  some phenomenological
understanding of this crossover phenomena  in the coarse grained level, a  microscopic picture
that consistently describes both the ballistic and the diffusive transport regimes has been lacking. 
In this work we  derive  a  scaling  from  for  the thermal  current   in a  class of one 
dimensional   systems  attached to   heat baths at boundaries,  
and show rigorously that the crossover occurs when the characteristic length scale of the system competes with the
system size.\end{abstract}

% {Nonelectronic thermal conduction and heat-pulse propagation in solids}

\maketitle

Low dimensional thermal transport \cite{Lebowitz, Lepiri, ADharRev} is a field of active 
research because of its surprises \cite{Liu2012,Yang2012} and intriguing features \cite{Hsiao2013,bae}. 
Most one dimensional (1D) models violate Fourier law and exhibit size dependent thermal conductivity
- the examples include harmonic chains \cite{ref8, ref9} the FPU models \cite{ref10, ref11} 
and the hard point gas model \cite{ref12, ref13, ref14}. Some 1D models with  
unusual scatterers \cite{LorenzGas,Triangle}, external potential \cite{DingLing, HardPointGas, AltMass} 
or strong nonlinearity  (e. g., interacting classical spins \cite{spin_sys, DB_PK}) however obey the 
Fourier law - the thermal current in 1D scales as $L^{-1}$ and the thermal conductivity
$\kappa (T)$ is an intensive constant dependent on the temperature and other system parameters. 
It has been seen that models which show diffusive heat propagation in the thermodynamic limit, 
also show a crossover to ballistic transport regime when system size is lowered \cite{DB_PK, DR_AD, PhononTransport}; 
the crossover scale is usually temperature dependent. This ballistic-diffusive crossover has also been
observed experimentally in many low-dimensional systems such as the graphene nanoribbon \cite{bae},
SiGe nanowires \cite{Hsiao2013}, stretched polymer nanofibres \cite{refB18}, carbon nanotubes \cite{refB17}
and other nanowires \cite{Nanowires}. %Details of some of the experimental observations and theoretical
%approaches are discussed in the supplemental material \cite{SM}. 
Several theoretical attempts have been made \cite{SM} to understand this ballistic-diffusive crossover in the  coarse grained level, 
dealing with Boltzmann transport equations \cite{Boltzman}, 
scaling theory \cite{DB_PK, Landi},  Langevin dynamics \cite{Langevin}, the Buttiker formalism \cite{taka}  or the 
nonequilibrium Green’s function approach \cite{jian} etc. However,  a unified microscopic formalism that goes beyond the recent phenomenological attempts 
\cite{Boltzman, Landi, Langevin, taka, jian, RevCasati, gilbert, daniel}
and consistently describes both the regimes is still lacking. Here we propose a scaling ansatz for the heat current in driven finite
systems in terms of the size and the characteristic length scale of the system. We rigorously derive this scaling form starting from
a microscopic dynamics and show that it  seamlessly connects the ballistic regime to the diffusive regimes.

Heat transfer in material systems generally involves a wide spectrum of energy carriers, e.g., electrons,
phonons, photons, molecules etc., but they operate at different energy and length scales - typically,
one of the energy carriers provides the dominant contribution to heat conductivity (see \cite{HeatCarrier}
for a comparison of length scales).
It is believed that scattering of the carriers, either at the material boundaries or from the impurities,
is essential for having normal heat transport.
In semiconductors or graphene nano-structures, heat carriers are predominantly phonons which are modeled
by the Boltzmann transport equation \cite{PhononTransport, Boltzman}. As such these studies are not only
of theoretical significance but also have immense technological implication \cite{RevCasati}. 

\begin{figure}
\centering
\includegraphics[width = 12.5cm]{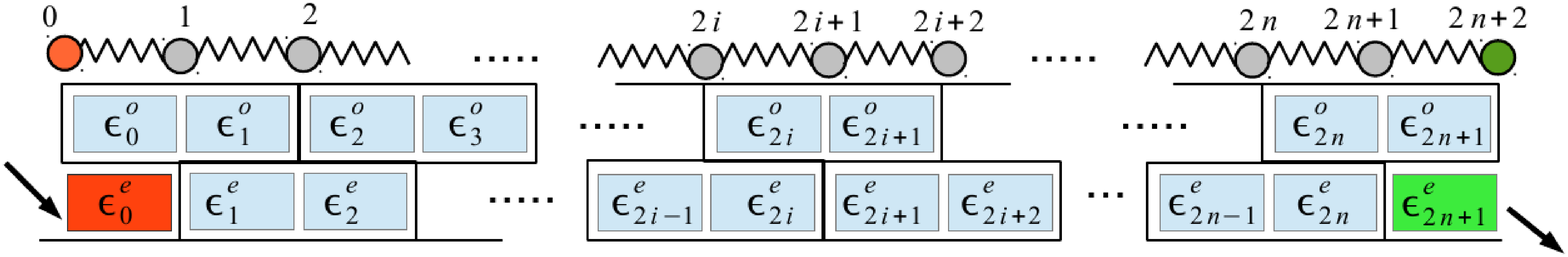}
\caption {Spring-mass model with parallel sub-lattice update:  
Each site of an 1D lattice ($i = 1,2,\dots, L=2n+1$  for system and $i=0,2n+2$ for bath) has 
a particle with position $x_i$. The  energy is given by  Eq. (\ref{H}). The first (second) row
corresponds to the odd (even) site update; boxes in each row depict the neighbouring pair
of bond energies $\eps_i^{o,e}$ which  are  exchanged  during the update.}
\label{fig:cartoon}
\end{figure}

In this article, we formulate a theoretical framework to study this ballistic-diffusive
crossover using a few simple 1D models. First we show that the microscopic energy conserving 
dynamics in these models can be described effectively by classical heat carriers. Next we show 
that any small scattering probability, either due to collision or interaction with other degrees
of freedom, results in diffusive behavior in the thermodynamics limit. Here the thermodynamic limit
implies that the system size is much larger than all other characteristic length scales present in
the system. The underlying idea of our work is that the characteristic length $\xi$ of the system
(mean free path of the  phonons or other  carriers,  or  the spin-spin correlation length in magnetic
systems) which is a function of average temperature of the system determines the ballistic to diffusive
crossover.
For $\xi \gg L$, the entire system is correlated (mean free path exceeds system size) and the carriers
propagate through the system without undergoing scattering and thus we have ballistic transport i.e.
$J \sim L^0$. In the other case, when $\xi \ll L$, the carriers undergo random scattering in the bulk of
the system and hence the transport is diffusive $J \sim L^{-1}$.

In solid insulators, thermal energy is predominantly transferred by phonons (lattice vibrations);
electron movements via collision appear in conductor. We start with a simple 1D model to describe
such lattice vibrations classically, which we refer to as the {\it coupled map lattice model}.
Consider a 1D lattice having particles with position variables $x_i$ ($i = 1,2...,L$). Let the
energy function of the model be defined as
\begin{equation}
E  = \frac k2 \sum_{i=1}^L (x_{i+1} - x_i)^2
\label{H}
\end{equation}
where $k$ is the coupling strength and $L = 2n+1$ $(n = 1,2,...)$. This is similar to the simple harmonic
oscillator but without the kinetic energy term. 

Using the transformation $u_i = \sqrt{k/2} (x_{i+1} - x_i)$ Eq. (\ref{H}) reduces to 
$E= \sum_i \epsilon_i$ where $\epsilon_i = u_i^2$ is the energy of the bond connecting
the masses $i$ and $i+1$. The heat bath is modeled by attaching two masses at the left and
right end with coordinates $x_0$ and $x_{L+1}$. These masses oscillate in such a way that
the bond energies $\epsilon_0$ and $\epsilon_L$ have Boltzmann distribution with temperatures
$\beta_0^{-1}$ and $\beta_L^{-1}$ respectively, i.e.,
\begin{equation}
P(\epsilon_0) = e^{-\beta_0 \epsilon_0} ~~ {\rm and ~~} P(\epsilon_L) = e^{-\beta_L \epsilon_L}.
\label{eq:e0eL}
\end{equation}

To study energy transport in this model we use a local odd-even parallel update scheme, where
lattice sites belonging to the even and odd sublattice are updated synchronously. 
To mimic the equation of motion, which conserves the total energy, we implement a dynamical
rule so that the energy is conserved locally. When a site $i$ is updated without disturbing
the neighbors $i\pm1$ (which belong to a different sublattice), both $u_i$ and $u_{i+1}$ are
updated keeping their sum $u_i+u_{i+1} = x_{i+1}-x_{i-1}$ unaltered. Along with this constraint,
energy conservation also demands that $u_i^2+u_{i+1}^2$ must be conserved. Thus the only possible
solutions are
\begin{eqnarray}
(u_i \to u_i, u_{i+1} \to u_{i+1})~ {\rm or}~( u_i \to u_{i+1}, u_{i+1} \to u_{i}).
\label{eq:backward}\nonumber
\end{eqnarray}
Clearly, the first solution does not update the positions $\{x_i\}$ and we chose to work with
the second  which effectively exchanges the energies $\epsilon_{i}$ and $\epsilon_{i+1}$ of the
bonds connected to the $i$-th site.
As can be seen from Fig. \ref{fig:cartoon}, during the odd sublattice update, the energies
$\epsilon^o_i$, of the bond that connect mass $i$ and $i+1,$ are exchanged pairwise:
$\epsilon^o_0 \rightleftharpoons \epsilon^o_1,
\epsilon^o_2 \rightleftharpoons\epsilon^o_2, \dots ,\epsilon^o_{2n} \rightleftharpoons \epsilon^o_{2n +1}$.
For the even sublattice update, energy exchange similarly occurs between
$\epsilon^e_1\rightleftharpoons\epsilon^e_2, \dots, \epsilon^e_{2n-1} \rightleftharpoons \epsilon^e_{2n}$ 
and the boundary energies $\epsilon^e_0$ and $\epsilon^e_L$ are refreshed by random values drawn
respectively from the Boltzmann distributions Eq. (\ref{eq:e0eL}).
Effectively, the energy which is introduced to the system at the site $i=0$ $(i=L)$ moves one step to the
right (left) during each sublattice update and goes out of the system through the other end exactly after
$L$ odd-even updates. This undeflected motion of the energy packets would result in ballistic heat-transport.
In the stationary state we have
\begin{equation}
\langle \epsilon^o_{2i} \rangle = \langle \epsilon_L \rangle~;~ \langle \epsilon^o_{2i+1} \rangle =\langle \epsilon_0 \rangle~;~
\langle \epsilon^e_{2i} \rangle = \langle \epsilon_0 \rangle~;~\langle \epsilon^e_{2i+1} \rangle = \langle \epsilon_L \rangle.\nonumber
\end{equation}
and thus the energy current $J = \langle \epsilon^e_{2i} \rangle - \langle \epsilon^o_{2i} \rangle$ is
independent of the system size. This result is same as the ballistic transport observed in the simple
harmonic lattice \cite{ref8} with usual Hamiltonian dynamics, the only difference being the absence of
the kinetic term here.

\begin{figure}
\centering
%\hskip-0.45cm
\includegraphics[width = 10.5cm]{cur.eps}\\\vspace*{.9 cm}
\includegraphics[width = 5.0cm]{random_seq.eps} 
\includegraphics[width = 5.6cm]{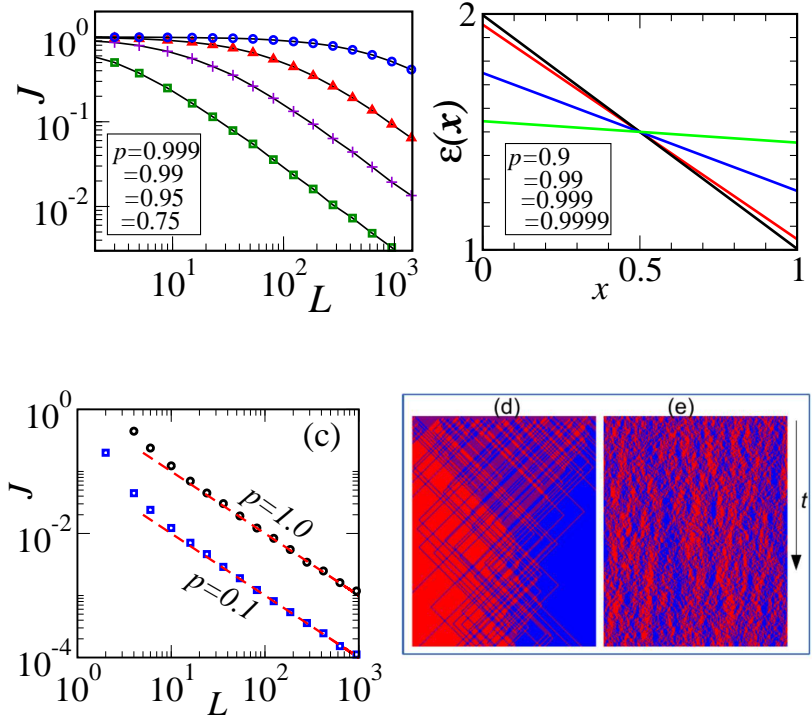}
\caption {{(a) Energy current $J$ versus $L$ (simulation results (symbols) are 
compared with Eq. (\ref{eq:J}) in solid line, and  (b) the energy profile $\eps(x)$ for 
different $p.$ Here $\langle \eps_0 \rangle = 2$, $\langle \eps_L \rangle = 1$ and ballistic 
limit is approached when $p\to1.$  
(c) $J$ versus $L$ for the model with random-sequential dynamics which results in diffusive
transport even when $p=1.$ (d) and (e)  show  evolution of the system for odd-even parallel  
and random-sequential dynamics respectively with $p=0.95$ and $L=200$. Energy carriers
introduced at the hot (left) and cold (right) boundaries are marked with different colors (red and blue).}}
\label{fig:Aplot}
\end{figure}

{\it Scattering of heat carriers:}
In order to have normal transport we need to introduce some scattering mechanism in this model, which
can change right movers (generated at the left end of the system) to left movers and vice versa. 
This can be incorporated if, during an update, particles change their position $x_i$ with a nonzero 
probability $p,$ i.e., now both the solutions of Eq. (\ref{eq:backward}) are chosen (the first one with
probability $1-p$ or otherwise the second one). Thus,
\begin{equation}
(\epsilon_i, \epsilon_{i+1}) \xrightarrow{1-p} (\epsilon_i, \epsilon_{i+1}); ~~
(\epsilon_i, \epsilon_{i+1})  \xrightarrow{p} (\epsilon_{i+1}, \epsilon_{i}).\nonumber
\end{equation}
This modified dynamics produces a scattering of both right and left movers with probability $q = 1-p$. Thus
the stationary energy profile must satisfy the following equations. 
\begin{eqnarray}
\epsilon^o_{2i} = q \epsilon^e_{2i} + p \epsilon^e_{2i+1}~~;~~ \epsilon^o_{2i+1} = q \epsilon^e_{2i+1} + p \epsilon^e_{2i};\cr
\epsilon^e_{2i} = q \epsilon^o_{2i} + p \epsilon^o_{2i-1}~~;~~\epsilon^e_{2i-1} = q \epsilon^o_{2i-1} + p \epsilon^o_{2i},
\end{eqnarray}
where $\epsilon^{o,e}_{i}$ are the average energy of the bond energy after odd and even sublattice updates.
In the second equation, which stands for even sublattice update, $i$ runs from $1$ to $n-1$ and the boundary
energies must satisfy 
$\epsilon^e_{0} = \langle \epsilon^e_0\rangle$, $\epsilon^e_{2n+1} = \langle \epsilon^e_L\rangle.$ 
This set of equations  can be solved
%can be written as 
% \begin{eqnarray}
% \epsilon^e_{2i} &=& (1 - 2 q) \epsilon^e_{2i-2} + 2 q \epsilon^e_{2i-1} \cr
% \epsilon^e_{2i+1} &=& - 2 q\epsilon^e_{2i-2} + (1+2 q) \epsilon^e_{2i-1}, \nonumber
% \end{eqnarray}
% which, 
along with the boundary conditions,  resulting in 
\begin{eqnarray}
\epsilon^e_{2i}  &=& (1 - 2 i q) \epsilon^e_0 + 2 i q \epsilon^e_1 \cr 
\epsilon^e_{2i+1} &=& - 2 i q \epsilon^e_0 + (1 + 2 i q) \epsilon^e_1.   
\end{eqnarray}
Thus the steady state energy profile    
$\langle \epsilon_i\rangle = \frac{1}{2} (\langle \epsilon^e_i\rangle  +\langle \epsilon^o_i\rangle)$ can 
be written, taking $x=i/L$ and $\Delta E=\langle \epsilon_0\rangle-\langle \epsilon_L\rangle,$ as
\begin{eqnarray}
%\epsilon(x)= \langle \epsilon_0 \rangle -\Delta E \frac{ p- 2Lqx}{2(Lq+p)}.
\epsilon(x)= \langle \epsilon_0 \rangle +\frac{p\Delta E }{2(Lq+p)} -\frac{ \Delta E}{(1+p/Lq)}x.
\end{eqnarray}
Here $L=2n+1$ is system size (excluding the two sites $i=0,L$ which are considered to be part of the heat reservoir).
Clearly  the profile  is linear, but  there are boundary layers 
at both ends (second term  in above equation) which vanish
when $L\to \infty.$ For any finite $L,$  however, the profile  becomes flat when $p\to 1$  (see Fig. \ref{fig:Aplot}(b)) -
an indication that energy
transport occurs ballistically  in this limit. Now the thermal current (Fig. \ref{fig:Aplot}(a)) is
\begin{eqnarray}
J &=&\epsilon^e_0 - \epsilon^o_0 = p (\epsilon^e_0 - \epsilon^e_1) 
= p \frac{\Delta E}{L q +p}.\label{eq:J} 
\end{eqnarray}

Thus, for any $p<1$, the coupled map lattice model shows diffusive transport
with conductivity $\kappa = p/q$ and is independent of system size.
The deterministic limit $p = 1$ however shows ballistic transport with heat current being
independent of the system size. Thus, diffusive  behaviour here is a consequence of stochasticity
which provides a scattering mechanism for the heat carriers, even though energy is conserved.
In fact if stochasticity is introduced differently, say by using a random sequential 
dynamics in this model instead of parallel sublattice update,  we  get diffusive
transport even for $p=1$ (see Fig. \ref{fig:Aplot}(c)).
To get a more physical picture of this ballistic-diffusive crossover let us rewrite
$J$ as
\begin{equation}
J = \kappa \frac{\Delta E}L {\cal F} \left(\frac \xi L \right) \label{eq:ansatz}
\end{equation}
where 
%${\Delta E}=\langle \epsilon_0\rangle-\langle \epsilon_L\rangle$ is the imposed energy gradient, 
$\xi= 1/q$ is the relevant length scale  (see discussions below) and ${\cal F}(x)$ is the scaling
function. In general ${\cal F}(x)$ has the following  property: ${\cal F}(0) = 1$ and ${\cal F}(x) \sim 1/x$
for large $x$ and for this simple model, ${\cal F}(x) = \frac1 {1 + p x}.$ In fact, this scaling form is not
accidental, we argue that such a scaling form always occurs in models of heat conduction and becomes prominent
only near the ballistic limit, where the dominant length scale of the system $\xi$ competes with the system
size $L$. This situation may arise near a critical point (where $\xi$ is the correlation length), in low
dimensional disordered systems ($\xi$ is the localization length), and in colliding particle systems ($\xi$ is
the mean-free path).

The model discussed here, can be interpreted as though energy carriers (say phonons) are generated
randomly and independently at both boundaries. At the left (right) end they are generated with unit
rate as right (left) movers to carry an energy unit $\epsilon_0 (\epsilon_L)$ drawn from a Boltzmann
distribution; they move in assigned directions (left or right) synchronously with unit rate and may
change their direction with probability $q$ (see Fig. \ref{fig:Aplot}(d) for evolution of the  
left and right movers). Thus they can move exactly $n$ steps persistently with probability
$q e^{-n/\xi}$ where $\xi= |\ln p|^{-1}$ is the persistent length which diverges in the limit
$q \to 0$. In this limit $\xi = |\ln(1-q)|^{-1} \sim q^{-1}$.

In a recent work \cite{DR_AD} a model of persistent walkers has been studied on a 1D lattice with
asynchronous (continuous time) update rules, where the right and left moving walkers can get interchanged at a
small rate $p$. These persistent walkers keep moving along one direction until they convert to the other type,
generating  a characteristic length scale similar to the mean free path of free carriers in a solid. This gives
rise to a length scale in the system. The authors showed that the electrical current can be expressed as
\begin{equation}
J_{el} = \frac{{\ell} v \Delta \rho}{L(1 + {\ell}/L)}
\label{DR_AD}
\end{equation}
where ${\ell}$ is the coherence length scale, $v$ being the mean velocity and $\Delta \rho$  the difference in  
carrier density  of left  and right reservoirs. Comparing this result with  Eq. (\ref{eq:J}), we find that in our
model $\ell \equiv p/q$ and $v = 1$ here. From purely phenomenological arguments, an expression similar to
Eqs. (\ref{eq:J}), (\ref{DR_AD})  energy transmission in carbon nanotubes \cite{jian,taka}, current in Heisenberg model \cite{DB_PK} has been introduced recently; a more  generic scaling form,  Eq. (\ref{eq:ansatz}), has also been  obtained for conductivity  in anharmonic chains \cite{Landi}.

\begin{figure}
\centerline{
\includegraphics[width=12.5cm]{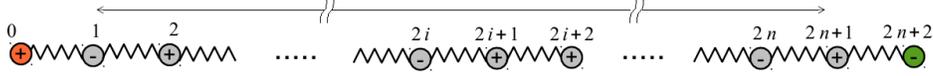}}
\caption{ Spring mass-spin model: Particle  at each site $i$ of the lattice 
($i = 1,2,\dots ,L=2n+1$  for system and $i=0,2n+2$ for bath)  has a spin $s_i = \pm$, 
along with the position variable $x_i,$ which evolve following Eq. (\ref{ising}). }
\label{fig:cartoon2}
\end{figure}

{\it Spring mass-spin model:}
Until now we have discussed models where heat carriers move independently. It is important to know if
Eq. (\ref{eq:ansatz}) still holds for a system in which the motion of heat carriers is correlated. 
To this end, we introduce the following model. Let us assume that every particle $i$ of the spring mass model
has a spin $s_i= \pm$. These Ising spins interact ferromagnetically, as described in Fig. \ref{fig:cartoon2}
and evolve following the Glauber dynamics
\begin{eqnarray}
\pm \xrightarrow{1 - \Delta V} \mp, ~ ~{\rm  where~} V = - \frac K 4 \sum_{i=0}^L s_i s_{i+1} - h s_0.
\label{ising}
\end{eqnarray}
The magnetic field $h$ at the left boundary controls the magnetization profile of the system. The
masses follow the parallel sublattice update rule with the energy function given by Eq. (\ref{ising}),
but now $x_i$ is updated (i.e., associated bond energies $\epsilon_i$ and $\epsilon_{i+1}$ are exchanged)
only when $s_i = +$. In this simple model, the spatial spin-spin correlation affects the energy transport
even though the spins do not take part directly and dictate only the update dynamics of the particles.

For simplicity, let us consider $h \to \infty$ limit, which forces the spin at site $i=0$ to be
$s_0= +$. %(for generic $h$, $s_0$ can also be $-$ with a nonzero probability).
The stationary probability $p_i$ that $s_i=+$ at site $i$, can be calculated 
from  the standard transfer matrix   $\langle s|T|s' \rangle = \exp(Kss'/4)$ as
\begin{eqnarray}
p_i&=& \frac 1 2 \left[1 + \tanh^i (K/4)\right].
\label{eq:pi}
\end{eqnarray}  
Now, $p_i$ is the scattering probability of the heat carriers at the lattice site $i$. In
other words, $p_i$ is the probability that a left (right) moving energy carrier becomes a
right (left) mover at site $i.$

We proceed with a generic function $p_i$, and will finally consider Eq. (\ref{eq:pi}) as a
special case.
Following the steps similar to the $p_i=p$ case, we find that in steady state, the average energies
after odd-sublattice update are 
\begin{eqnarray}
\epsilon^o_{2i} = q_{2i+1} \epsilon^e_{2i} + p_{2i+1}  \epsilon^e_{2i+1}; ~
\epsilon^o_{2i+1} =  q_{2i+1} \epsilon^e_{2i+1} + p_{2i+1}  \epsilon^e_{2i}, \nonumber
\end{eqnarray} 
and the same after even-sublattice update ($i=1,2,  \dots, n-1$) are 
\begin{eqnarray}
\epsilon^e_{2i} =  q_{2i} \epsilon^o_{2i} + p_{2i}  \epsilon^o_{2i-1} ;~
\epsilon^e_{2i-1} = q_{2i} \epsilon^o_{2i-1} + p_{2i}  \epsilon^o_{2i},\nonumber
\end{eqnarray}
where $q_i = 1-p_i.$ These equations need to be solved with boundary conditions 
$\epsilon^e_{0}= \langle \epsilon_0 \rangle$  and $\epsilon^e_{L}= \langle \epsilon_L \rangle.$
It is now straightforward  to obtain the 
energy profile (after odd-sublattice update),
\begin{eqnarray}\hspace*{-.5 cm}
\epsilon^e_{2i} &=&(1-\lambda_i) \langle \epsilon_0\rangle + \lambda_i \epsilon^e_1,~
\epsilon^e_{2i+1} =  \epsilon^e_{2i} +  p_1 \frac{\langle \epsilon_0\rangle -   \epsilon^e_1}{p_{2i+1}}\cr
&& {\rm with} ~ \lambda_i=  p_1\sum_{j=1}^{2i +1} p_j^{-1} -2 p_1 i.
\label{prof}
\end{eqnarray}

To calculate the energy profile explicitly we need $\epsilon^e_1$; 
the boundary condition $\epsilon^e_{2n+1}=\langle \epsilon_L \rangle$ gives 
\begin{eqnarray}
\epsilon^e_1= \frac{(\lambda_n-1)\langle\epsilon_0\rangle +  \langle \epsilon_L \rangle }{\lambda_n}= 
\langle\epsilon_0\rangle- \frac{\Delta E}{\lambda_n}.
\label{eq:e1}
\end{eqnarray}
It is evident, from Eqs. (\ref{eq:J}) and (\ref{eq:e1}), that  the energy current is
\begin{eqnarray}
J =  p_1  \frac{ \Delta E }{ \lambda_n}
\simeq \frac{ \Delta E }{L(\alpha_1-1) +1}
\label{JJ}
\end{eqnarray}
where  in the last step we have taken the continuum  limit  $x=i/L$  to write  Eq. (\ref{prof}) as,
\begin{equation} \lambda_n \simeq p_1  [ L(\alpha_1 -1) +1] ~~{\rm  with }~~
%\alpha_x  = \int_0^x \frac{dy}{p(y)}. \label{eq:alpha}
\frac{d\alpha_x}{dx}  = \frac{1}{p(x)} \label{eq:alpha}
\end{equation}
In the continuum limit, the energy profile (from (\ref{prof}))  is 
\begin{eqnarray}
 \epsilon(x) = \epsilon_0 +\frac{\Delta E}{2L(\alpha_1-1) +1} -  \Delta E \frac{L(\alpha_x-1) +1}{L(\alpha_1-1) +1}. \nonumber
\end{eqnarray}
Note,   that  for a given   boundary drive, both current  and  energy profile  depend 
{\it only on} the  scattering probability $p(x)$  (through $\alpha_x$ from Eq. (\ref{eq:alpha})); 
details of the dynamics, interaction or disorder plays a role in generating a  spatial 
variation  of scattering.  
These exact results,  though  derived here for  a class  of interacting systems  driven at boundaries, are   expected to be  generic;
it explains   in general, why  current and  density profile  in   driven systems depend only   on the  
scattering probability  of  the current-carriers  in the bulk of the sample.

Evidently the current $J$  in Eq. (\ref{JJ}) becomes  ballistic when $\alpha_1 = 1$ and for any $\alpha_1 >1$
one gets the Fourier law in the thermodynamic limit $L \gg (\alpha_1-1)^{-1}.$
Thus $(\alpha_1-1)^{-1}$ plays the role of correlation length $\xi,$ which diverges in the
limit $\alpha_1 \to 1.$ To verify  that $(\alpha_1-1)$ actually plays the role of $\xi^{-1},$
let us consider a simple example $p(x) =  e^{-\gamma x}$ where $\gamma^{-1}$ is a measure of
spatial correlation. Now $\alpha_1-1 =\gamma^{-1} (e^{\gamma} -1 -\gamma)$ which, for small
$\gamma,$ gives $\gamma/2.$ Thus  $J= \Delta E / (1+ L\gamma/2)$ is consistent with the ansatz
(\ref{eq:ansatz}). Again, for a spring mass-spin model $p_i$ is given by Eq. (\ref{eq:pi});
then $\alpha_1-1 \simeq |\ln(\tanh(K/4))|/4$, which is inversely proportional to the correlation
length known for Ising model. Also note that for $p_i = p$ we trivially recover the result
Eq. (\ref{eq:J}) from Eq. (\ref{JJ}).

{\it  Discussion:}
The model that we have introduced here can be considered to be formally identical to a system
of free particles with momentum ${p_i}$, $H = \sum_i p_i^2/2m$. The odd-even dynamics is nothing but the
integration of the equation of motion using a finite integration time step $\Delta t$. The
additional constraint introduced in spring mass system, i.e., $u_i + u_{i+1}$ is not altered
during update of $x_i$, translates to momentum conservation during collision of particles $i$
and $i+1$. Finally, the scattering probability $p$ in spring mass system becomes the collision
probability. Thus one obtains ballistic transport when collision occurs pairwise and deterministically,
whereas energy transport is diffusive when collisions are missed out  with finite probability.
Using these simple models we have proven rigorously that the scaling ansatz proposed for the 
current, Eq. (\ref{eq:ansatz}) can seamlessly describe both the ballistic transport regime with 
the diffusive regime. There are however a few drawbacks of the current formalism. Firstly, the heat 
carriers are assumed to carry a quantized packet of energy with them which does not get scattered 
and redistributed, and therefore unrealistic for real life transport problems. Secondly in  the spring 
mass-spin model, introduced as a model with spatial correlation, the dynamics of the background 
spins is completely decoupled from that of the heat carriers. It remains a challenge to address 
these issues in the future.

 {\it Acknowledgments:} PKM  would   like to  acknowledge the support of CEFIPRA  under Project 4604-3.

\end{document}